\documentclass[preprintnumbers,letter,twocolumn,aps,nofootinbib]{revtex4}

\usepackage{epsfig}
\usepackage{amsmath}
\usepackage{amssymb}
\usepackage{verbatim}
\usepackage{bm}
\usepackage{dsfont}
\usepackage[footnotesize]{caption}
\usepackage{subfigure}
\usepackage{cancel}

\def\r{\rho}
\def\rb{\bar{\rho}}
\def\Sb{\bar{S}}
\def\p{\phi}
\def\pb{\bar{\phi}}
\def\s{\sigma_0}
\def\a{\alpha}
\def\ab{\bar{\alpha}}
\def\d2{W''}

\begin{document}

\begin{flushleft}
DESY 14-031\\
April 2014
\end{flushleft}

\title{Supersymmetric Moduli Stabilization and High-Scale Inflation}

\author{Wilfried Buchmuller}
\affiliation{Deutsches Elektronen-Synchrotron DESY, 22607 Hamburg, Germany}
\author{Clemens Wieck}
\affiliation{Deutsches Elektronen-Synchrotron DESY, 22607 Hamburg, Germany}
\author{Martin Wolfgang Winkler}
\affiliation{Deutsches Elektronen-Synchrotron DESY, 22607 Hamburg, Germany}

\begin{abstract}
We study the back-reaction of moduli fields on the inflaton potential in generic models of F-term inflation. We derive the moduli corrections as a power series in the ratio of Hubble scale and modulus mass. The general result is illustrated with two examples, hybrid inflation and chaotic inflation. We find that in both cases the decoupling of moduli dynamics and inflation requires moduli masses close to the scale of grand unification. For smaller moduli masses the CMB observables are strongly affected.
\end{abstract}

\maketitle

%
%%
%%%
%%%%
%%%%%%%%%%%%%%%%%%%%%%%%%%%%%%%%%%%%%%%%%%%%%%%
%%%%%%%%%%%%%%%%%%%%%%%%%%%%%%%%%%%%%%%%%%%%%%%
%%%%%%%%%%%%%%				 %%%%%%%%%%%%%%%%%%%%%%%
%%%%%%%%%%%%%	%         Introduction    %%%%%%%%%%%%%%%%%%%%%%%
%%%%%%%%%%%%%%				  %%%%%%%%%%%%%%%%%%%%%%
%%%%%%%%%%%%%%%%%%%%%%%%%%%%%%%%%%%%%%%%%%%%%%%
%%%%%%%%%%%%%%%%%%%%%%%%%%%%%%%%%%%%%%%%%%%%%%%

\section{Introduction}

Supergravity F-term inflation is an attractive theoretical framework for describing the anisotropies of the cosmic microwave background radiation \cite{Ade:2013uln}. This includes hybrid inflation \cite{Copeland:1994vg,Dvali:1994ms} as well as chaotic inflation \cite{Kawasaki:2000yn} for which a large class of models has recently been constructed \cite{Kallosh:2010xz}.

Supergravity is a low-energy effective theory, which motivates attempts to embed supergravity inflation into higher-dimensional UV-complete theories, in particular string theory. However, in such constructions stabilization of all moduli including the dilaton, K\"ahler and complex structure moduli is a potential problem. A standard procedure is to use gaugino condensates and fluxes \cite{Carlos:1992da,Giddings:2001yu,Kachru:2003aw}, which can lead to metastable Minkowski vacua. In general, moduli stabilization in the entire cosmological history leads to upper bounds on the reheating temperature \cite{Buchmuller:2004xr} and the energy scale of inflation \cite{Kallosh:2004yh}. In the simple setup by KKLT the modulus mass and the barrier protecting the vacuum are proportional to the gravitino mass. This leads to a tension between high-scale inflation and low-energy supersymmetry breaking.

As a possible alternative we study supersymmetric moduli stabilization, in the sense that all moduli masses are large and independent of the gravitino mass. Examples of this type are `racetrack models'  where the height of the barrier protecting metastable Minkowski space can be arbitrarily large, independent of the scale of supersymmetry breaking \cite{Kallosh:2004yh}. Detailed analyses of moduli dynamics in specific inflation and supersymmetry breaking models have been carried out in \cite{Davis:2008sa,Davis:2008fv,Kallosh:2011qk,Buchmuller:2013uta}. 

In this Letter we consider the consistency of supersymmetric moduli stabilization and generic F-term inflation. In general, the coupling to the modulus generates corrections to the inflaton potential which can be expanded in powers of $H/m_\rho$, the ratio of Hubble scale during inflation and modulus mass.

In small-field inflation, such as hybrid inflation, this produces a linear term in the inflaton at leading order. This is analogous to the effect of supersymmetry breaking which induces a linear term proportional to the gravitino mass \cite{Buchmuller:2000zm}. Depending on its size such a linear term can have a significant effect on inflationary observables, in particular the spectral index of scalar fluctuations \cite{Nakayama:2010xf,Pallis:2013dxa,Buchmuller:2013dja,Buchmuller:2014epa}. 

In chaotic inflation the leading order correction is suppressed by an additional power of $H/m_\rho$ compared to hybrid inflation. Nevertheless, the modulus-induced terms can have severe consequences for CMB observables in the case of a large Hubble scale during inflation, which is suggested by
the recently released B-mode polarization data \cite{Ade:2014xna}.

%
%%
%%%
%%%%
%%%%%%%%%%%%%%%%%%%%%%%%%%%%%%%%%%%%%%%%%%%%%%%
%%%%%%%%%%%%%%%%%%%%%%%%%%%%%%%%%%%%%%%%%%%%%%%
%%%%%%%%%%%%%%				 %%%%%%%%%%%%%%%%%%%%%%%
%%%%%%%%%%%%%	%         Section 2        %%%%%%%%%%%%%%%%%%%%%%%
%%%%%%%%%%%%%%				  %%%%%%%%%%%%%%%%%%%%%%
%%%%%%%%%%%%%%%%%%%%%%%%%%%%%%%%%%%%%%%%%%%%%%%
%%%%%%%%%%%%%%%%%%%%%%%%%%%%%%%%%%%%%%%%%%%%%%%

\section{Supersymmetric moduli stabilization and inflation}

In a four-dimensional effective theory a single modulus $\rho$ residual from a higher-dimensional supergravity theory is usually described by the classical tree-level K\"ahler potential\footnote{We restrict our discussion to a single modulus, assuming that all other moduli are stabilized supersymmetrically at a higher scale.}
\begin{align}
K_{\rm mod}(\r,\rb) = -\kappa \log{(\r + \rb)}\,,
\end{align}
where, for example, $\kappa = 1$ for the dilaton in heterotic string theory and $\kappa = 3$ for a K\"ahler modulus in type IIB string theory. Stabilization of $\rho$ is achieved by an appropriate superpotential $W_\text{mod}(\rho)$. In the remainder of this Letter we assume that $W_\text{mod}$ is such that the scalar potential has a local minimum at $\r_0 = \rb_0 \equiv \s$ which is supersymmetric and Minkowski, i.e.,
\begin{align}\label{eq:Modconditions}
W_{\rm mod}(\s) = 0 \, , \quad D_\r W_{\rm mod}(\s) = 0,
\end{align}
where $D_\rho$ denotes the K\"ahler covariant derivative with respect to $\rho$. If the conditions eq.~\eqref{eq:Modconditions} are met one finds
\begin{align}
m_\r = \frac{1}{\kappa} (2\s)^{2-\kappa/2} \,W_\text{mod}''(\sigma_0)\,,
\end{align}
where primes denote derivatives with respect to $\rho$. To ensure the modulus remains stabilized its mass is required to be large compared to the scale of any other dynamics in the effective theory, for example, the scale of inflation.

An example for this kind of moduli stabilization, the KL-model, was developed in \cite{Kallosh:2004yh}. Here the superpotential is of a racetrack type, i.e.,
\begin{align}
W_{\rm mod}(\r) = W_0 + A e^{-a\r} + B e^{-b\r}\,,
\end{align}
and $W_0$ is tuned such that $W_\text{mod}(\sigma_0)$ vanishes. The metastable Minkowski vacuum at $\rho = \sigma_0$ is separated from a global AdS minimum by a high barrier . In addition, $m_\rho$ is independent of the gravitino mass in the effective theory, regardless of the mechanism which breaks supersymmetry.

To study the impact of a stabilized modulus with a large supersymmetric mass on models of inflation in supergravity, we consider a theory defined by 
\begin{align}
K_{\rm tot} &= -\kappa \log{(\r + \rb)} + K(\p_\a, \pb_{\ab})\,,\nonumber \\ \quad W &= W_{\rm mod}(\r) + W_{\rm inf}(\p_\a) \,,
\end{align}
where $\phi_\alpha$, with $\alpha = 1,2, ...$, denote the chiral superfields in the inflation model. Hence, we study the most general theory without explicit couplings between the two sectors. The scalar potential can be written as 
\begin{align}\label{eq:treepot}
V = \frac{e^{K}}{(\r + \rb)^\kappa}\bigg\{&\frac{(\r +
    \rb)^2}{\kappa} |D_\r W|^2 \nonumber \\
    &+ K^{\a\ab}D_{\a}W  D_{\ab} \overline{W} - 3 |W|^2\bigg\}\,.
\end{align}
Generically, there is a non-trivial interaction between the modulus and inflaton sectors. Due to the large positive energy density during inflation the minimum of the modulus is shifted by $\delta \rho$. The displacement is obtained by imposing $\partial_\rho V|_{\s+\delta \rho}=0$ at the new minimum. The expression for $\delta \rho$ can be expanded in powers of $H/m_\rho$, where $H = \sqrt{V/3}$ is the Hubble scale during inflation. The aforementioned requirement of $m_\rho > H$ makes this analysis self-consistent. Including all terms up to second order in $H/m_\rho$, we find\footnote{We assume that $W_\text{mod}''$ is not hierarchically suppressed compared to higher derivatives of $W_\text{mod}$.}
\begin{align}
\delta \rho = &\; \frac{W_\text{inf}}{(2\s)^{\kappa/2-1} \,m_\rho} + \frac{1}{(2\s)^{\kappa-1}\, m_\rho^2} \bigg[K^{\alpha \bar \alpha} D_\alpha W_\text{inf} \partial_{\bar \alpha} \overline W_\text{inf}\nonumber \\ 
&- |W_\text{inf}|^2 - W_\text{inf}^2 \left( 1-\frac{\kappa}{2} + \frac{W_\text{mod}'''(\sigma_0)}{2 \kappa \, (2 \sigma_0)^{\kappa/2-3} \, m_\rho} \right)\bigg]  \nonumber \\ 
&+ \mathcal{O}\left(\frac{H^3}{{m_\rho}^3}\right)\,.
\end{align}
This implies
\begin{align}
D_\r W|_{\s+\delta \rho} = &\;\frac{\kappa}{(2\s)^{\kappa/2+1}\,  m_\rho} K^{\a\ab} D_{\a} W \partial_{\ab}
\overline{W} \nonumber \\
&+ \mathcal{O}\left(\frac{H^2}{{m_\rho}^2}\right)\,,
\end{align}
i.e., $D_\r W$ is suppressed by one power of $m_\rho$. After setting the modulus to its proper minimum, the inflaton potential reads
\begin{align}\label{eq:effpot} 
V= &\;\;\frac{V_\text{inf}(\phi_\alpha)}{(2 \sigma_0)^\kappa}\nonumber \\ \nonumber
&- \frac{\kappa}{2 (2\s)^{3\kappa/2} \,m_\rho} \Big\{ W_\text{inf} \Big[V_\text{inf}(\phi_\alpha) \nonumberÊ\\
&+ e^K K^{\alpha \bar \alpha} \partial_\alpha W_\text{inf} D_{\bar \alpha} \overline W_\text{inf} \Big] + \text{ c.c.} \Big\}\nonumber \\
&- \frac{\kappa \, e^K}{(2\s)^{2\kappa}\, m_\rho^2} \Big|K^{\alpha \bar \alpha} D_\alpha W_\text{inf} \partial_{\bar \alpha} \overline W_\text{inf} \Big|^2\,,
\end{align}
at leading order in $H/m_\rho$. Here $V_\text{inf}(\phi_\alpha)$ denotes the inflationary potential in the absence of a modulus sector, i.e.,
\begin{align}
V_\text{inf}(\phi_\alpha) = e^K \Big\{ K^{\alpha \bar \alpha} D_\alpha W_\text{inf} D_{\bar \alpha} \overline W_\text{inf} - 3|W_\text{inf}|^2 \Big\}\,.
\end{align}
Notice that all powers of $2\sigma_0$ in eq.~\eqref{eq:effpot} can be absorbed by a redefinition of $W_\text{inf}$.
As naively expected, all corrections vanish in the limit of an infinitely heavy modulus. In the following, we study the effect of the leading order terms in two representative examples of F-term inflation.

%
%%
%%%
%%%%
%%%%%%%%%%%%%%%%%%%%%%%%%%%%%%%%%%%%%%%%%%%%%%%
%%%%%%%%%%%%%%%%%%%%%%%%%%%%%%%%%%%%%%%%%%%%%%%
%%%%%%%%%%%%%%				 %%%%%%%%%%%%%%%%%%%%%%%
%%%%%%%%%%%%%	%         Section 3        %%%%%%%%%%%%%%%%%%%%%%%
%%%%%%%%%%%%%%				  %%%%%%%%%%%%%%%%%%%%%%
%%%%%%%%%%%%%%%%%%%%%%%%%%%%%%%%%%%%%%%%%%%%%%%
%%%%%%%%%%%%%%%%%%%%%%%%%%%%%%%%%%%%%%%%%%%%%%%

\section{Examples}

%
%%
%%%
%%%%
%%%%%%%%%%%%%%%%%%%%%%%%%%%%%%%%%%%%%%%%%%%%%%%
%%%%%%%%%%%%%%%%%%%%%%%%%%%%%%%%%%%%%%%%%%%%%%%
\subsection*{Hybrid Inflation}
As a first example we consider F-term hybrid inflation with a canonical K\"ahler potential \cite{Dvali:1994ms},
\begin{align}
K = \p\pb + S_1\bar{S}_1 + S_2\bar{S}_2\,,
\end{align}
where $\p$ and $S_{1,2}$ denote the inflaton and waterfall fields, respectively. As is well known, with this form of $K$ the $\eta$-problem of supergravity inflation is evaded for a superpotential linear in $\p$, in particular for
\begin{align}
W= \lambda \p \left(\frac{v^2}{2} - S_1 S_2\right)\,,
\end{align}
which defines the hybrid inflation model. During inflation, as long as $\phi > \phi_\text{c} = v/\sqrt{2}$, the waterfall fields remain at the origin of field space, and the superpotential effectively reduces to
\begin{align}
W_{ \rm inf}= \frac{1}{2}\lambda v^2 \p \,.
\end{align}
The corresponding scalar potential reads
\begin{equation}\label{eq:onlyHI}
V=\frac{1}{4}\lambda^2 v^4 + V^{\rm loop} + V^{\rm sugra}\,.
\end{equation}
In the small field regime the slope of the potential is dominated by the Coleman-Weinberg potential $V^{\rm loop}$ while the supergravity corrections $V^{\rm sugra}$ can be neglected. The latter contain only quartic and higher order terms in $\phi$. Notice that the potential eq.~\eqref{eq:onlyHI} only depends on $|\p|$. 

Including the leading order modulus correction eq.~\eqref{eq:effpot}, the effective tree-level potential is given by
\begin{align}\label{eq:withLT}
 V_{\rm tree}(\p)= V_0\, \left\{1- \frac{\,\sqrt{V_0}}{m_\rho}\,\kappa\,(\p+\pb) \right\} +\mathcal{O}\left(\p^3\right)\,,
\end{align}
with
\begin{align}
 V_0 = \frac{1}{4}\tilde{\lambda}^2\,v^4\,,\qquad \tilde{\lambda}^2 = \frac{\lambda^2}{(2\s)^\kappa}\,.
\end{align}
Evidently, the modulus induces a linear term in the inflaton potential. In the limit $m_\rho\rightarrow \infty$ the original potential is recovered up to a total rescaling factor which can be absorbed by a redefinition of $\lambda$. 

A linear term in the inflaton field is also induced by soft supersymmetry breaking \cite{Buchmuller:2000zm}. Depending on its size relative to the one-loop potential, it can significantly affect the dynamics of inflation \cite{Buchmuller:2000zm,Nakayama:2010xf,Pallis:2013dxa}. This is important in particular for the spectral index of scalar fluctuations, which in hybrid inflation is typically $n_s \simeq 0.98$ \cite{Dvali:1994ms}. This value can be reduced to the currently measured value $n_s \simeq 0.96$ if the linear term is taken into account \cite{Pallis:2013dxa,Buchmuller:2013dja,Buchmuller:2014epa}.

Based on eq.~\eqref{eq:withLT} one may expect that the modulus alters inflation for 
$m_\rho \lesssim \sqrt{V_0} \sim H$. But, as we will show, important corrections to the CMB observables already arise at considerably larger values of $m_\rho$. The linear term in the inflaton potential distinguishes the real and imaginary part of $\p$. Therefore, it effectively turns the single-field inflation model into a two-field model which has recently been studied in detail \cite{Buchmuller:2014epa}. A qualitative understanding of the constraints on the modulus mass can be obtained by considering the special case of inflation along the real axis.

In this case the canonically normalized inflaton is $\varphi=(\p+\pb)/\sqrt{2}$. The modulus affects the slope of the inflaton potential: including the one-loop potential we find
\begin{equation}\label{eq:slope}
\frac{V'(\varphi)}{V(\varphi)}\simeq \frac{\tilde{\lambda}^2}{8\pi^2}\,\frac{1}{\varphi} - \frac{\kappa}{\sqrt{2}}\,\frac{\tilde{\lambda}\,v^2}{m_\rho}\,,
\end{equation}
where we have assumed $\varphi_\text{c}\ll \varphi \ll 1$. The relative sign of the two contributions depends on the sign of $\varphi$ during inflation. A particularly interesting situation arises when the modulus-induced slope partially cancels the slope of the Coleman-Weinberg potential. This flattens the inflationary trajectory and reduces the distance in field space corresponding to the $N_* \sim 55$ e-folds of inflation. More specifically, $\varphi_*$, the value of $\varphi$ at horizon crossing of a scale relevant for the CMB observables, is reduced. At smaller $\varphi_*$ the potential is more curved, i.e., the second slow roll parameter $\eta$ increases, which in turn affects the spectral index.  For a careful choice of the inflationary parameters and $m_\rho$, a value $n_s\simeq 0.96$ can be obtained. Hence, modulus-induced corrections to the inflaton potential can be used to reconcile F-term hybrid inflation with Planck observations. 

Notice that the linear term induces a second minimum in the inflaton potential. Therefore, a careful choice of initial conditions is necessary to avoid that $\varphi$ gets trapped in the second minimum. However, this problem is alleviated when considering inflation in the full complex plane rather than along the real axis~\cite{Buchmuller:2014epa}. In this case a wide range of initial conditions leads to successful inflation in accordance with the Planck data.

To estimate the value of $m_\rho$ at which the modulus corrections become relevant, let us first consider hybrid inflation without the modulus. Neglecting supergravity corrections and assuming $\varphi_*\gg \varphi_\text{c}$, one finds ${\varphi_*=\tilde\lambda\sqrt{N_*}/2\pi}$. Furthermore, with $N_*\sim 55$, the measured amplitude of scalar fluctuations implies ${v\simeq 0.3M_{\rm GUT}}$ \cite{Dvali:1994ms}, where $M_{\rm GUT} \simeq 2\cdot 10^{16}\,\text{GeV}$ is the scale of gauge coupling unification in the supersymmetric standard model. By use of eq.~\eqref{eq:slope} we then find that the modulus induces an $\mathcal{O}(1)$ correction to the slope of the potential if
\begin{align}\label{eq:criticalmodmass}
 m_\rho \sim  2\pi \sqrt{2N_*} \kappa v^2 \sim 6\cdot 10^{-2} \kappa M_{\rm GUT}\,.
\end{align}
This implies that even if the modulus is stabilized at a scale close to $M_{\rm GUT}$, it can significantly affect the dynamics of hybrid inflation. Notice that $m_\rho$ in eq.~\eqref{eq:criticalmodmass} is much larger than the naive estimate $m_\rho \sim H \sim 10^{11}\,\mathrm{GeV}$, where we have assumed $\tilde{\lambda} \sim 10^{-2}$, the largest coupling for which hybrid inflation works.

In \cite{Buchmuller:2014epa} a detailed analysis has been carried out for hybrid inflation with a linear term whose strength is controlled by the gravitino mass, ${V_{3/2} = -\lambda v^2 m_{3/2} (\phi + \bar \phi)}$. Comparing this expression with eq.~\eqref{eq:withLT}, one can match the parameters which yield successful inflation with ${n_s \simeq 0.96}$. Let us consider an example with large coupling, $\lambda \simeq 3\cdot 10^{-3}$, for which the condition $\varphi_* \gg \varphi_\text{c}$ is fulfilled. The measured spectral index is then obtained\footnote{The spectral index $n_s \simeq 0.96$ is related to $v \simeq 4\cdot 10^{15}\,\mathrm{GeV}$ and $m_{3/2} \simeq 10^5\,\mathrm{GeV}$, see Fig.~3 in\cite{Buchmuller:2014epa}. This corresponds to the modulus mass $m_\rho \mathrel{\widehat=} \kappa\lambda^2 v^4/8m_{3/2}$.} for $m_\rho \simeq 6\cdot 10^{-2} \kappa M_{\rm GUT}$, in agreement with the estimate eq.~\eqref{eq:criticalmodmass}. For smaller couplings $\lambda$ the modulus mass which gives the right spectral index decreases.

We conclude that in hybrid inflation modulus corrections are generically important even if the modulus is stabilized close to the GUT scale. This may appear surprising since the Hubble parameter is much smaller than $M_\text{GUT}$, but it is a consequence of the enormous flatness of the inflaton potential.

%
%%
%%%
%%%%
%%%%%%%%%%%%%%%%%%%%%%%%%%%%%%%%%%%%%%%%%%%%%%%
%%%%%%%%%%%%%%%%%%%%%%%%%%%%%%%%%%%%%%%%%%%%%%%
\subsection*{Chaotic Inflation}
As a second example we consider chaotic inflation. Its simplest supergravity embedding is defined by the K\"ahler potential \cite{Kawasaki:2000yn,Kallosh:2010xz}
\begin{align}
K = -\frac{1}{2}(\p-\pb)^2 + S\Sb - \xi (S\Sb)^2\,,
\end{align}
and the superpotential
\begin{align}
W_{ \rm inf} = S f(\p)\,.
\end{align}
The scalar field $S$ generates the inflaton potential via its F-term but decouples from the inflationary dynamics. This situation is engineered by including a sufficiently large $(S\Sb)^2\text{-term}$ in the K\"ahler potential. The latter lifts the mass $m_S$ beyond the Hubble scale during inflation and stabilizes $S$ at the origin of field space. The inflaton is identified with the real part of $\p$ which is protected against supergravity corrections by a shift symmetry.
Including the leading order modulus correction, the scalar potential can be expressed as
\begin{align}
e^{(\p-\pb)^2/2} V = &\, \,m_S^2 |S|^2 + |\tilde{f}(\p)|^2 + \mathcal{O}(|S|^4)\nonumber\\
&-\frac{\kappa\,|\tilde{f}(\p)|^2}{m_\rho} \Bigg[ \left(S\tilde{f}(\p)+ \text{ c.c.}\right) \nonumber \\                                       		&+ \frac{|\tilde{f}(\p)|^2}{m_\rho} + \mathcal{O}(|S|^2)\Bigg]\,,
\end{align}
with
\begin{align}
\tilde{f}(\p) &= \frac{f(\p)}{(2 \sigma_0)^{\kappa/2}}\,, \nonumber \\ m_S^2 &=  \left|\tilde{f}'(\p)-(\p-\pb)\,\tilde{f}(\p)\right|^2+4\xi \left|\tilde{f}(\p)\right|^2\,.
\end{align}
Notice that the modulus induces a displacement of $S$ which in turn affects the potential of $\phi$. At leading order in $H/m_\rho$, the new minimum lies at
\begin{align}
 \bar{S}=\frac{\kappa \,|\tilde{f}(\p)|^2\,\tilde{f}(\p)}{m_\rho\, m_S^2}\,.
\end{align}
The inflaton potential at the minimum of $S$ reads
\begin{align}
e^{(\p-\pb)^2/2} \;V(\phi) &= V_0(\phi) \left\{ 1 - \kappa\frac{V_0(\phi)}{m_\rho^2} - \kappa^2\frac{V_0^2(\phi)}{m_\rho^2\,m_S^2}  \right\}\,,\nonumber \\
 V_0(\phi) &= \left|\tilde{f} (\phi )\right|^2\,.
\end{align}
In case the imaginary part of $\p$ is stabilized at the origin, the exponential factor becomes unity and $V_0$ denotes the inflaton potential without modulus corrections. This happens for standard choices of $f(\p)$, in particular for monomial functions. Furthermore, notice that the leading modulus correction only appears at order $m_\rho^{-2}$. The absence of a correction of order $m_\rho^{-1}$ results from the suppression of $W_\text{inf}$ by one power of $m_\rho$. This is an important difference compared to the case of hybrid inflation.

In chaotic inflation the scale $V_0(\phi_*)$ is fixed by the amplitude of scalar fluctuations, without modulus corrections one finds $V^{1/4}_0(\phi_*)\simeq  M_{\rm GUT}$. Thus, our perturbative analysis breaks down for $m_\rho \lesssim M_{\rm GUT}^2$ and inflation and modulus stabilization are no longer decoupled. For modulus masses up to $M_{\rm GUT}$ significant corrections to the CMB observables arise.

%
%%
%%%
%%%%
%%%%%%%%%%%%%%%%%%%%%%%%%%%%%%%%%%%%%%%%%%%%%%%
%%%%%%%%%%%%%%%%%%%%%%%%%%%%%%%%%%%%%%%%%%%%%%%
%%%%%%%%%%%%%%				 %%%%%%%%%%%%%%%%%%%%%%%
%%%%%%%%%%%%%	%         Discussion     %%%%%%%%%%%%%%%%%%%%%%%
%%%%%%%%%%%%%%				  %%%%%%%%%%%%%%%%%%%%%%
%%%%%%%%%%%%%%%%%%%%%%%%%%%%%%%%%%%%%%%%%%%%%%%
%%%%%%%%%%%%%%%%%%%%%%%%%%%%%%%%%%%%%%%%%%%%%%%

\section{Conclusion}

We have analyzed the back-reaction of a supersymmetrically stabilized modulus on F-term inflation in supergravity. Generically, the inflaton potential receives corrections due to a shift of the modulus minimum which can be written as a power series in the ratio of Hubble scale during inflation and modulus mass. Hence, in the limiting case of an infinitely heavy modulus all corrections vanish. For a modulus mass between the Hubble and the Planck scale there can be sizeable effects in many F-term inflation models, as we have demonstrated in two examples. 

In hybrid inflation the leading order correction is linear in the inflaton and necessitates a two-field description. For a modulus mass close to the GUT scale, the correction term can resolve the well-known tension between the predicted and the measured scalar spectral index. In fact, one can show that any small-field inflation model with non-vanishing superpotential receives a leading order linear correction, with a wide and model-dependent range of possible effects. Note, however, that small-field inflation models are disfavored by the recently released BICEP2 data \cite{Ade:2014xna}.

In chaotic inflation the leading order correction to the scalar potential is suppressed by an additional power of $H/m_\rho$ compared to hybrid inflation. This is due to the fact that the superpotential is suppressed itself. Nevertheless, the leading correction may have similarly grave effects on predicted observables as in hybrid inflation.

The BICEP2 data suggest a large Hubble parameter during inflation, $H \sim M^2_{\rm GUT}\sim 10^{14}\,\mathrm{GeV}$. According to our analysis this has severe implications for higher-dimensional theories. Stabilized extra dimensions during inflation require moduli masses close to the scale of grand unification, $M_{\rm GUT}$, which appears to coincide with the energy scale of inflation.

%
%
%
%%
%%%
%%%%
%%%%%%%%%%%%%%%%%%%%%%%%%%%%%%%%%%%%%%%%%%%%%%%
%%%%%%%%%%%%%%%%%%%%%%%%%%%%%%%%%%%%%%%%%%%%%%%
\subsection*{Acknowledgments}
This work has been supported by the German Science Foundation (DFG) within the Collaborative Research Center 676 ``Particles, Strings and the Early Universe''. The work of C.W. is supported by a scholarship of the Joachim Herz Stiftung.

%
%%
%%%
%%%%
%%%%%%%%%%%%%%%%%%%%%%%%%%%%%%%%%%%%%%%%%%%%%%%
%%%%%%%%%%%%%%%%%%%%%%%%%%%%%%%%%%%%%%%%%%%%%%%
%%%%%%%%%%%%%%				 %%%%%%%%%%%%%%%%%%%%%%%
%%%%%%%%%%%%%	%         Bibliography   %%%%%%%%%%%%%%%%%%%%%%%
%%%%%%%%%%%%%%				  %%%%%%%%%%%%%%%%%%%%%%%
%%%%%%%%%%%%%%%%%%%%%%%%%%%%%%%%%%%%%%%%%%%%%%%
%%%%%%%%%%%%%%%%%%%%%%%%%%%%%%%%%%%%%%%%%%%%%%%

\end{document}